\documentclass{JINST}
\usepackage{epsfig}
\usepackage{textcomp}

\title{Fabrication and Characterisation of Oil-Free Large High Pressure Laminate Resistive Plate Chamber}

\author{Rajesh Ganai$^a$\thanks{Corresponding author.}, Arindam Roy$^a$, Kshitij Agarwal$^b$, Zubayer Ahammed$^a$, Subikash Choudhury$^a$
and Subhasis Chattopadhyay$^a$\\
\llap{$^a$}Variable Energy Cyclotron Centre,\\
  1/AF-Bidhan Nagar, Kolkata-700064, India\\
\llap{$^b$}Birla Institute of Technology and Science,\\
  Pilani, Rajasthan-333031, India\\
  E-mail: \email{rajesh.ganai.physics@gmail.com}}

\abstract{A large (240 cm $\times$ 120 cm $\times$ 0.2 cm) oil-free High Pressure Laminate (HPL), commonly referred as
``bakelite'', Resistive Plate Chamber (RPC) 
has been developed at VECC-Kolkata using locally available P-302 OLTC grade HPL. 
The chamber has been operated in streamer mode using Argon, Freon(R134a) and Iso-butane in a ratio of 34:57:9 by
volume. The electrodes and glue samples have been characterised by measuring their electrical parameters
like bulk resistivity and surface resistivity. The performance of the chamber has been studied by measuring the efficiency, its
uniformity and stability in detection of cosmic muons. Timing  measurement has been performed at a central location of the chamber.
The chamber showed an efficiency $>$95$\%$ and time resolution ($\sigma$), at the point of measurement, $\sim$0.83 ns at 9000V. 
Details of the material characterisation, fabrication procedure and performance studies have been 
discussed.}


\begin{document}

\section{Introduction}

Resistive Plate Chamber (RPC) \cite{1} is a type of gas-filled detector that utilises a constant and uniform electric field produced 
between two highly resistive ($10^{9}$ $\Omega$cm - $10^{13}$ $\Omega$cm) parallel electrodes made of materials like glass or 
High Pressure Laminate (HPL), commonly referred as ``bakelite''. 
Relatively low cost, large surface area and very good time resolution ($\sim$0.5 ns) \cite{1} make RPC suitable for triggering
and detection of muons in several high energy experiments like CMS \cite{2}, ATLAS \cite{3}, BELLE-II \cite{4}, BABAR \cite{5},
BES-III \cite{6}. 
The Iron CALorimeter (ICAL) experiment in the India based Neutrino Observatory (INO) \cite{7} and the Near 
Detector (ND) of the Deep Underground Neutrino Experiment (DUNE) at Fermilab, USA \cite{8} are two upcoming neutrino experiments that will use RPCs for detection of muons. 
ICAL of dimension $\sim$ 48 m $\times$ 16 m $\times$ 14 m will consist of $\sim$ 50kT magnetised iron plates stacked in 150 layers \cite{7}.  
RPC modules each of dimension $\sim$ 200 cm $\times$ 200 cm $\times$ 0.2 cm sandwiched between two iron plates will be used 
as tracking layers. The RPC modules to be used in  DUNE are of dimension 200 cm $\times$ 100 cm $\times$ 0.2 cm. 
The present work is aimed at developing RPCs for these two experiments. 

In this paper, we have discussed the fabrication and characterisation of a (240 cm $\times$ 120 cm $\times$ 0.2 cm) RPC 
that uses 3 mm thick HPL sheets as electrodes. 

Among the experiments world-wide which use HPL RPCs, most of them use electrodes coated with linseed oil.
The physical features like roughness, defects  on the uncoated inner surface of 
the RPC electrodes may cause high leakage current, high noise rate \cite{9} which may result into breakdown of the electrodes 
\cite{10}. Coating the inner surface of the HPL sheets with  polymerised oil like linseed has been 
a common practice for ensuring the long term stability of the RPC modules. It has been seen that a thin layer of linseed \cite{11} 
or silicone oil \cite{12} coated on the inner surface of the HPL electrodes can significantly
improve the surface smoothness, thereby greatly improving the performance of the RPCs. 

Application of uniform coating on the electrodes adds several complexities in the fabrication procedure. Additionally, the 
adverse effects of coating on the properties of electrodes are completely eliminated with the use of uncoated surface.
Also, surface treatment of the electrodes with oil has its own disadvantages which are well documented in \cite{13}.  
In the case of oil-treated RPCs, uncured oil droplets in the form of "stalagmites" \cite{13} have been observed on the inner surface of the HPL plates. These droplets offer
a suitable path to the current through the gas gap leading to high leakage current. The chance of 
accumulation of these droplets is very high around the spacers of the chamber. It has also been observed that the surface
resistivity of the oil-treated HPL changes during its course of operation. 
These problems have been however, solved by the use of minimal, cured linseed oil \cite{14}.
This work is a parallel effort towards constructing a large-sized HPL RPC without any surface treatment.
The glossy finished electrode surfaces have not been further treated
with any lubricants like linseed oil, silicone oil for smoothness.
Since 1990's several R\&Ds have been done to develop HPL RPCs without any kind of oil
treatment on its surface\cite{15}, \cite{16}. These R\&Ds were mainly focussed to improve the surface quality of the HPL sheets
by using fine paper and melamine resins \cite{16}. However, even after that, the noise rate of the uncoated RPCs were considerably higher
compared to the coated RPCs \cite{17}. Hence, in most of the cases it has been noted that the attempts to eliminate the 
oiling did not give satisfactory results mainly due to significant increase in the noise level of the RPCs \cite{17}, \cite{18}.

In the next section, the characterisation of the ingredients like
electrodes, glue have been discussed. The details of fabrication procedure and the cosmic ray test set-up have been discussed in 
sections 3 and 4 respectively. The results have been discussed in section 5 followed by concluding remarks in section 6.

\section{Characterisation of electrodes and glue}

\subsection{Electrical properties of the HPL sample}
\begin{figure}
\begin{center}
\includegraphics[height=5.0 cm, width=8.0 cm,keepaspectratio]{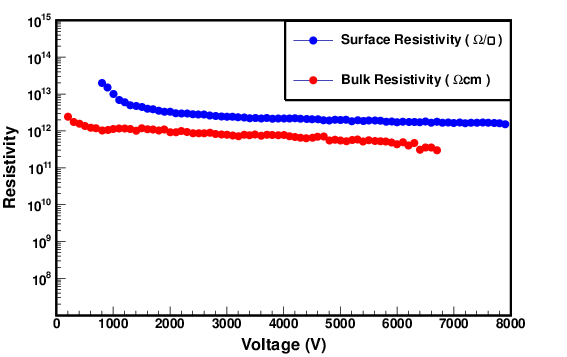}
\caption{\label{setup2} \small \sl [Color online] Electrical properties of the HPL sample as a function of the applied voltage}
\label{HPL-rest}
\end{center}
\end{figure}

The electrical properties like bulk and surface resistivities of the electrodes are 
important parameters \cite{19},\cite{20} that decide the suitability of the electrodes in fabricating a chamber.
High resistivity controls the rate capability of the chamber and helps to localize the avalanche. The average thickness of the electrodes used here is 3 mm.
The resistivities have been measured by a specially designed jig in which the sample is inserted in between 
two copper sheets connected
to the opposite terminals of the power supply. Fig.~\ref{HPL-rest} shows the measured bulk and surface 
resistivities of the HPL sample as a function of the
applied high voltage. The average value of the bulk resistivity of the HPL sample was 
found to be $\sim$ 9 $\times$ $10^{11}$ $\Omega$cm whereas the 
surface resistivity was measured to be $\sim$ 3 $\times$ $10^{12}$ $\Omega$/$\Box$.
The values are found to suit the requirements of RPC electrodes \cite{21}. 

\subsection{Electrical properties of different glue samples}
The glue applied on top of the spacers while fabricating the chamber primarily provides required mechanical 
strength at the joints, however, the electrical properties of the glue plays an important role in deciding the chamber
properties.
Since the glue contributes to the leakage current of the RPC, the conductivity of the glue used should be much less than 
that of the electrodes.

\begin{figure}
\begin{center}
\includegraphics[height=5.0 cm, width=8.0 cm,keepaspectratio]{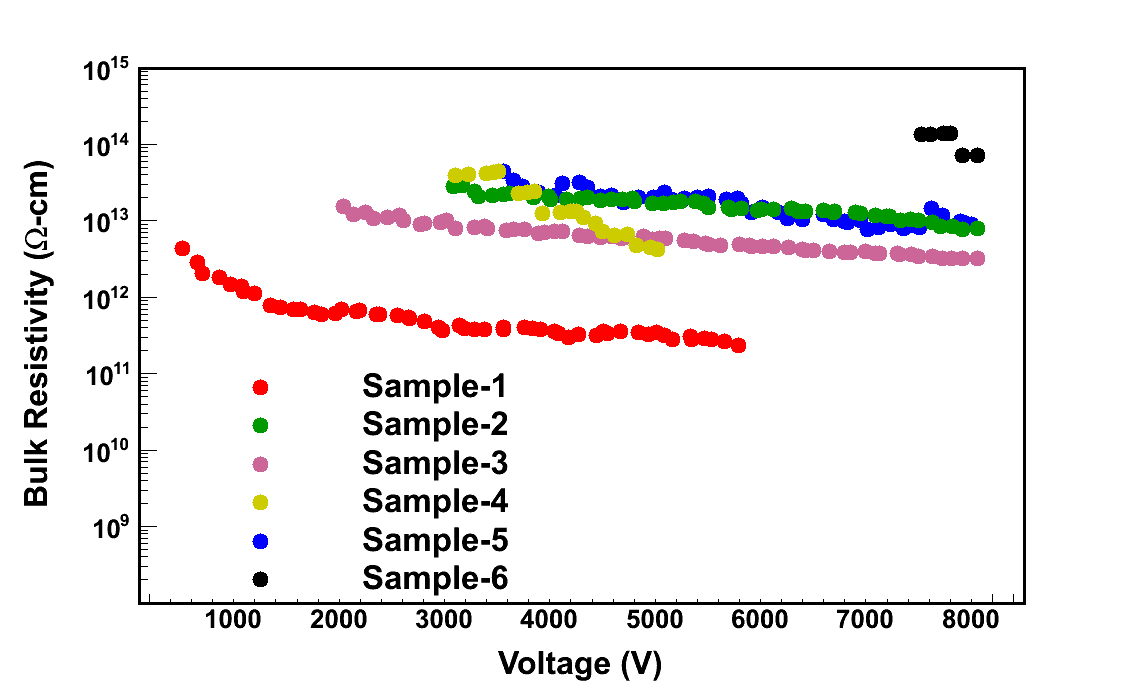}
\caption{\small \sl [Color online] Bulk resistivity of different glue samples as a function of the applied voltage.}
\label{glue}
\end{center}
\end{figure}

Table-1 shows the resin and hardener specifications used in preparing six different glue samples. 
The bulk resistivity of the glue samples have been measured. Fig.~\ref{glue} shows the variation of the bulk resistivity 
of the samples with the applied voltage. It is seen that the resistivity of most of the glue samples are higher 
compared to that of the electrodes.
 
\begin{table}
\begin{center}		
\begin{tabular}{|c|c|c|}
\hline
Glue sample    & Resin Specifications	& Hardener Specifications \\
\hline
Sample-1 &    Dobekot 520F	&	Hardener 758        \\
\hline
Sample-2 &	Araldite	&      Araldite hardener      \\
\hline
Sample-3 & 	Dobekot 520F  &		Hardener 758	\\
\hline
Sample-4 &      Dobekot 520F	&	Fevitite hardener\\
\hline
Sample-5 &	 Bicron BC-600	&	Hardener 758\\
\hline
Sample-6   &     BC-600:Araldite::1:1 	&	BC-600 hardener\\
\hline
\end{tabular}
\caption[]{\label{tab:nucl} \small \sl Resin and hardener specifications of different glue samples.}
\end{center}		
\end{table}

Table-2 summarizes the mixing proportions (by mass) of the resin and hardener for
different glue samples and their respective bulk resistivities. The bulk 
resistivity of Sample-6 ($\sim$ $10^{14}$ $\Omega$cm) has been found to be $\sim$100 
times higher than that of the HPL electrode ($\sim$ $10^{12}$ $\Omega$cm), hence this particular glue has been used to fabricate 
the chamber.

\begin{table}
\begin{center}		
\begin{tabular}{|c|c|c|}
\hline
Glue sample    &Resin:Hardener (by mass)	& Average $\rho$ ($\Omega$cm) \\
\hline
Sample-1	&	 1.0 : 0.8 &    $\sim$ 6.773 $\times$ $10^{11}$        \\
\hline
Sample-2	&	 1.0 : 1.0 &    $\sim$ 2.164 $\times$ $10^{13}$         \\
\hline
Sample-3	&	 11.0 : 1.0 &   $\sim$ 8.376 $\times$ $10^{12}$\\
\hline
Sample-4	&	 21.0 : 2.0 &   $\sim$ 2.014 $\times$ $10^{13}$   \\
\hline
Sample-5	&	 1 : 1 &$\sim$ 6.62 $\times$ $10^{12}$ \\
\hline
Sample-6	&	 4:1  &      $\sim$ 1.157 $\times$ $10^{14}$ \\
\hline
\end{tabular}
\caption[]{\label{tab:nucl} \small \sl Mixing ratio and bulk resistivity ($\rho$) of different glue samples.}
\end{center}		
\end{table}

\section{Fabrication of the chamber}
 
The fabrication of such a large-sized oil-free chamber has several challenges which include

(a) maintaining the planarity of such large-sized electrodes
(b) uniformity in coating the surface with semi-conducting paint 
(c) preventing the electrodes from sagging by using spacers at proper locations
(d) ensuring continuous and uniform gas flow through the detector
(e) proper sealing of the chamber to ensure the detector is gas-tight.
In this section how these challenges were overcome during the fabrication procedure have been discussed.
The fabrication procedure consisted of the following steps 
(a) erection of a suitable assembly platform
(b) filing the edges and chamfering the corners of the electrodes
(c) cleaning of the electrodes
(d) painting of the electrodes with semi-conducting paint 
(e) measurement of levelling of the electrodes 
(f) pasting the spacers on the lower electrode and application of glue 
(g) installation of the upper electrode.
For the procedure of oil coating the elecrodes, a step is likely to be added in between (c) and (d). For uniform oil coating 
of such a large surface, a specialized zig was to be made and specialized coating procedure required to be adopted. This becomes
an additional step whcih we avoided by not using any kind of oil.
During the fabrication of the RPC, the electrode sheets have been kept on a well levelled platform.
 In order to maintain a constant gas gap, we  used two types of spacers- side spacers and button
spacers. These spacers have been fixed on the HPL sheets with the help of the chosen glue. 
The button spacers helped to maintain the gas gap while the side spacers additionally helped to seal the chamber from
all sides.
For such a large-sized RPC, the spacers served the important purpose of providing mechanical support alongwith 
defining the gas gap.
The surface area of the spacers in contact with the electrodes should be adequate 
enough to provide excellent mechanical strength to the RPC. The option was to increase either the number or the surface area of
the button spacers. Initially, the number of button spacers was increased which led to the
accumulation of gas in certain regions of the chamber. The enhanced gas pressure in these regions led to the popping out of the
button spacers. Therefore, it has been decided to increase the area of each button spacer. A similar problem, bulging of the 
chamber due to accumulation of gas was seen initially when there were two gas inlets and two gas outlets. A change in the number
of gas inlets and outlets from two to four helped to solve the problem.

In the discussions to follow, some of the steps of fabrication have been described in detail.
 
\subsection{Levelling measurement}
\begin{figure}
\begin{center}
\includegraphics[height=4.0 cm, width=7.0 cm]{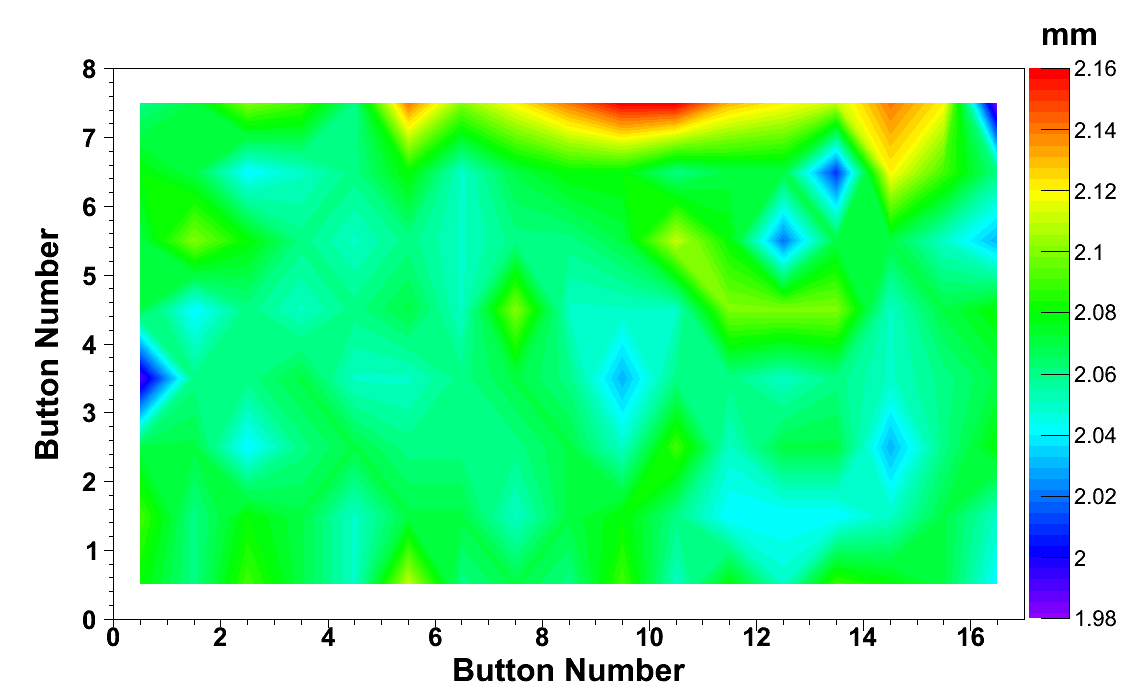}
\caption{\label{setup2} \small \sl [Color online] Variation of local height (mm) of glue and button spacers pasted 
on the lower HPL sheet.}
\label{local-height-glued-button}
\end{center}
\end{figure}

In order to fabricate such a large RPC, we need a platform of good planarity and of comparable dimensions as that of
the chamber. A good, plane platform
ensures that the HPL sheets do not sag and the spacers stick properly onto both the electrodes.
We built a special platform placing cardboard sheets, foam and a thick (2 cm) glass plate of dimensions
$\sim$(240 cm $\times$ 120 cm) on top of each other. These components ensured a well levelled surface for the assembly 
of the chamber.
128 different locations were marked on the bottom electrode in the form a (16 $\times$ 8)
matrix and the local heights of the electrode at those positions were measured with the help of a dial gauge indicator. 
After gluing the button spacers on those 128 locations, the local heights of the glued buttons pasted on 
the electrode were measured, the distribution of which is illustrated in
Fig.~\ref{local-height-glued-button}. It is seen from 
Fig.~\ref{local-height-glued-button} that the gap thickness remains uniform.

\subsection{Assembly of the chamber}

As a first step of the preparation of the electrodes, all the edges of both 
the HPL sheets were filed properly for smoothening. All the
surfaces of both the sheets were then properly cleaned with de-mineralised water and alcohol. After cleaning the sheets, 
one surface of each
sheet was spray-painted with a black semi-conducting paint mixed in the ratio 1:1 by volume with a special dry 
thinner, both manufactured by Kansai Nerolac, India. The resistance profile of the painted surfaces was measured
with the help of a jig made of two brass rods of 9 cm length, separated by a distance of 9 cm from each other. 
Fig.~\ref{rest-profile-paint-1} and 
Fig.~\ref{rest-profile-paint-2} show the surface resistivity profile of the painted surfaces of the two electrodes. 

\begin{figure}[!h]
\begin{center}
\includegraphics[height=4.0 cm, width=7.0 cm]{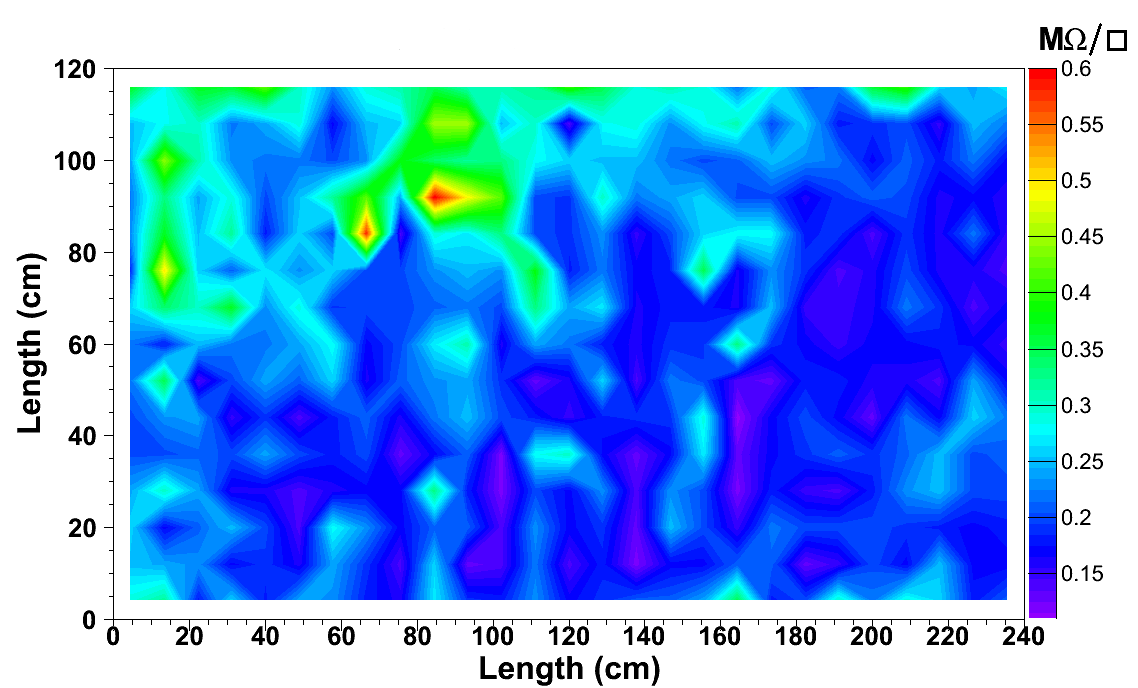}
\caption{\label{rest-profile-paint-1} \small \sl [Color online] Surface resistivity profile of the painted  surface of the lower electrode.}
\label{rest-profile-paint-1} 
\end{center}
\end{figure}

\begin{figure}[!h]
\begin{center}
\includegraphics[height=4.0 cm, width=7.0 cm]{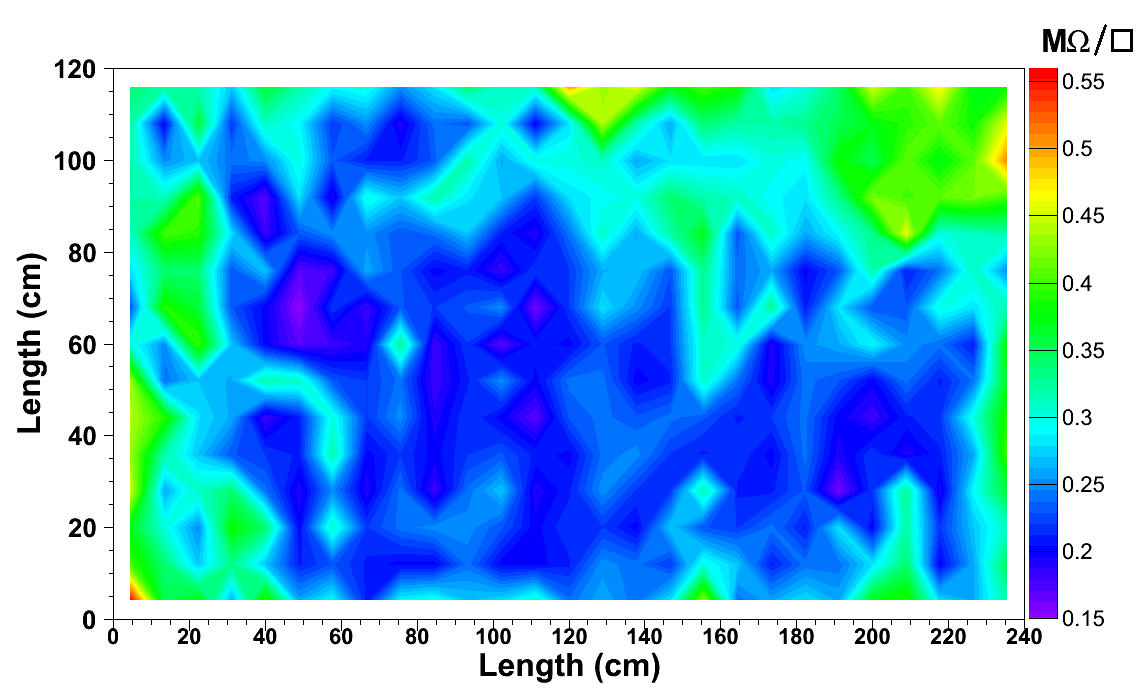}
\caption{\label{rest-profile-paint-2} \small \sl [Color online] Surface resistivity profile of the painted surface of the upper electrode.}
\label{rest-profile-paint-2} 
\end{center}
\end{figure}

Fig.~\ref{rest-proj-1} 
and Fig.~\ref{rest-proj-2} show the uniformity of painting on the two surfaces in terms of the distribution of the measured 
resistivities. Even though the RMS values of the two distributions (31$\%$ and 25$\%$) indicate relatively larger variation, 
the deviations are mostly at the edges where the paints are relatively non-uniform at the end of the spray-gun runs. The
RMS widths are $\sim$5$\%$ and $\sim$6$\%$ after excluding the tails. The applied field is therefore expected to be uniform.

\begin{figure}
\begin{center}
\includegraphics[height=5.0 cm, width=8.0 cm]{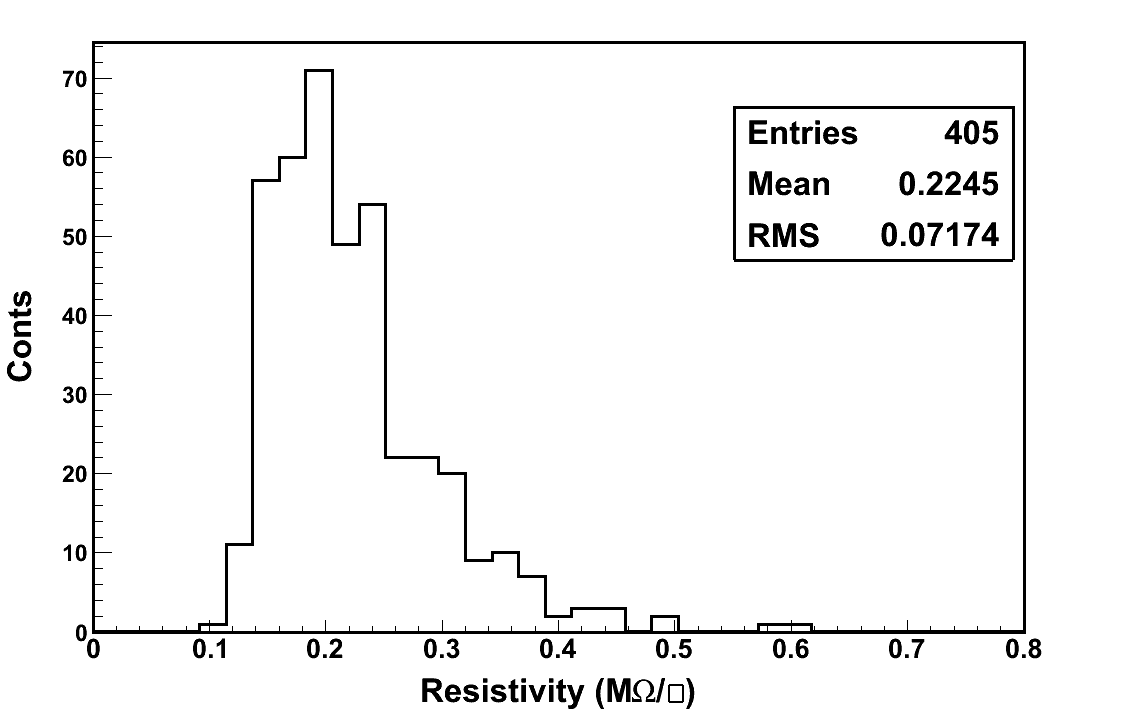}
\caption{\label{rest-proj-1} \small \sl Surface resistivity distribution of the lower HPL surface.}
\label{rest-proj-1} 
\end{center}
\end{figure}

\begin{figure}
\begin{center}
\includegraphics[height=5.0 cm, width=8.0 cm]{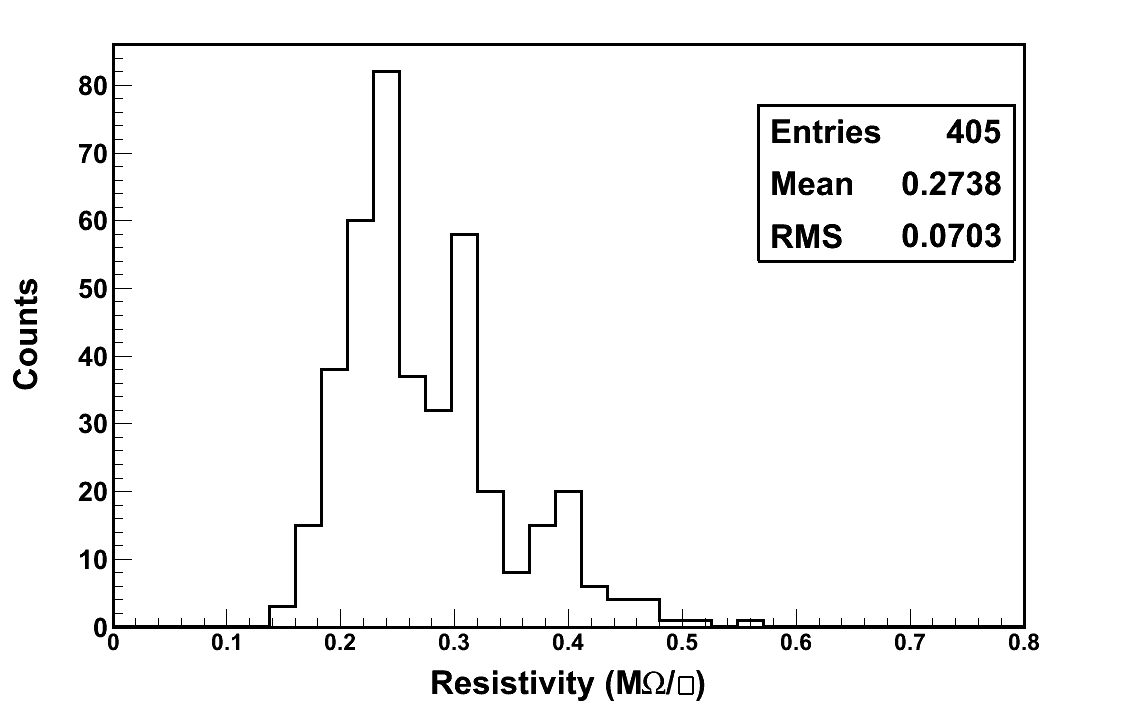}
\caption{\label{rest-proj-2} \small \sl Surface resistivity distribution of the upper HPL surface.}
\label{rest-proj-2} 
\end{center}
\end{figure}

\begin{figure}
\begin{center}
\includegraphics[height=3.6 cm, width=6.0 cm]{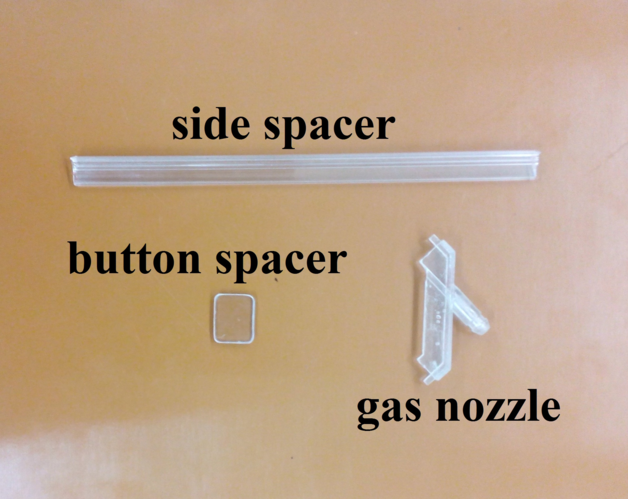}
\caption{\label{setup2} \small \sl [Color online] Components (not to scale) used in the fabrication of large HPL RPC. The side spacer is a sample 
piece of the large side spacers used.}
\label{spacers}
\end{center}
\end{figure}

Two copper tapes each of dimension (16 cm $\times$ 2.5 cm) were pasted at the edges of the painted surfaces of the electrodes.
The tapes are used to apply high voltages on the surfaces. The painted surfaces were then isolated properly with mylar 
sheets and kapton tapes.  
The side spacers, button spacers and the gas nozzles were glued subsequently. A total of 128 button
spacers each of size 1.5 cm $\times$ 1 cm , 6 side spacers each of $\sim$80 cm in length, 8 gas nozzles 
(4 for gas input and 4 for gas output) and 2 side spacers each of $\sim$120 cm in length have been used. The distance
between any two button spacer is $\sim$14 cm. Some of the components have been illustrated in Fig.~\ref{spacers}.

\begin{figure}
\begin{center}
\includegraphics[height=4.2 cm, width=5.6 cm]{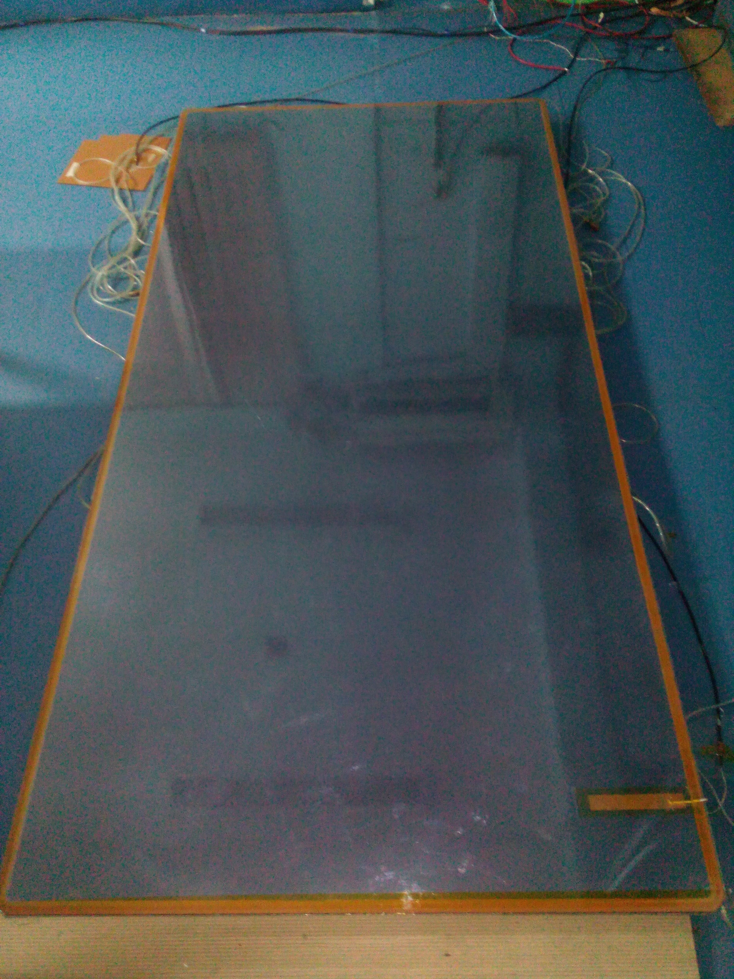}
\caption{\label{setup2} \small \sl [Color online] Photograph of the large HPL RPC.}
\label{full-rpc}
\end{center}
\end{figure}

\begin{figure}
\begin{center}
\includegraphics[height=4.0 cm, width=7.0 cm,keepaspectratio]{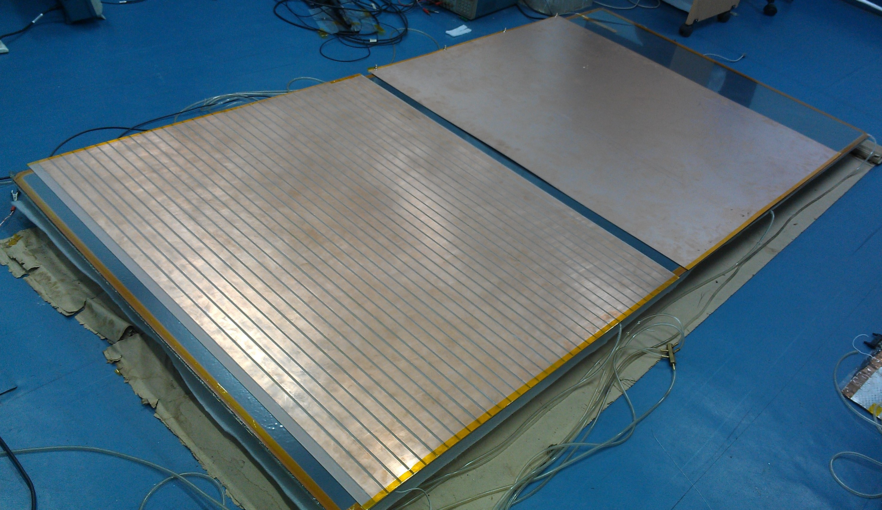}
\caption{\label{setup2} \small \sl [Color online] Photograph of the large HPL RPC with pick-up panel.}
\label{full-rpc-with-pickup}
\end{center}
\end{figure}

The upper electrode was installed after the application of glue on top of all the spacers. A number of weights (12) each of 
$\sim$ 2.350 kg, placed over the mylar surface on the top electrode and kept for one day to ensure better clinging.
We then reglued the
side spacers to ensure gas-tightness of the chamber. The chamber was then ready for testing with gas and High Voltage. 
Fig.~\ref{full-rpc} shows the photograph of the complete RPC and Fig.~\ref{full-rpc-with-pickup} shows the complete RPC 
with the pick-up panels on it.
The pick-up panels are made of $\sim$(125 cm $\times$ 105 cm $\times$ 0.15 cm) FR4 sheet sandwiched between 
$\sim$(125 cm $\times$ 105 cm $\times$ 0.0035 cm) copper sheets. The copper pick-up strips are 2.5 cm in width, with a gap of 0.2 cm 
between adjacent strips. 

\section{Cosmic ray Test set-up}

The RPC has been tested with cosmic rays in a standard cosmic ray test set-up. We have used three plastic scintillators - two
paddle scintillators (20 cm $\times$ 8.5 cm) and one finger scintillator (7 cm $\times$ 1.5 cm). 
The overlap area between the scintillators has been used to obtain 
the cosmic ray efficiency for a particular set-up.

High voltage (HV) was applied to the chamber using the CAEN A1832PE and A1832NE modules in the CAEN SY1527 crate.  
The current was monitored from the panel of the HV supply. The signal from the chamber
was tapped with the help of LEMO connectors soldered on the copper strips of the pick-up panel. 
A CANBERRA QUAD CFD 454 constant fraction
discriminator (CFD) has been used to digitise the signals from the scintillators and the RPC. 
The coincidence of the three scintillator logic signals form the 3-fold master trigger. Further coincidence with the RPC 
logic signal forms the four-fold signal. The efficiency is defined as the ratio of the 4-fold counts to the 3-fold counts 
during a fixed time interval. For timing measurements, the master trigger i.e. 3-fold
was connected to the TDC-START and the RPC logic signal was sent to the TDC-STOP after a fixed delay.
A CAMAC based data acquisition system has been used in our setup.
The average master trigger rate was $\sim$ 0.008 Hz/$cm^2$.

\section{Test results and discussions}
During the entire testing period, the laboratory temperature
has been maintained at $\sim$ 20\textdegree C  and the relative humidity has been maintained at $\sim$45\% - 55\%. 
All the tests have been done in the streamer
mode of operation of the RPC with a gas composition of Argon:Freon(R134a):Iso-butane::34:57:9 by volume. 
A typical gas flow rate of 
$\sim$0.75 litre/hour has been maintained over the entire test period resulting in $\sim$3 changes
of gas volume per day.
The current of the detector has remained stable over the period of $\sim$120 days during which the chamber 
has remained in operation.

\subsection{I-V characteristics}

The I-V characteristics of the fabricated HPL RPC is shown in Fig.~\ref{I-V}. Two distinct slopes in the I-V characteristics 
have been obtained with a breakdown voltage at $\sim$7000V. From the Ohmic part, the calculated bulk resistivity of the 
chamber was found to be ~1.72$\times$$10^{13}$ $\Omega$cm. Fig.~\ref{bulk_with_time} shows the variaton of bulk resistivity of the chamber 
over a period of $\sim$ a month. The variation of temperature and relative humidity during the measurements are shown in 
Fig.~\ref{efficiency_time}.
\begin{figure}
\begin{center}
\includegraphics[height=5.0 cm, width=8.0 cm]{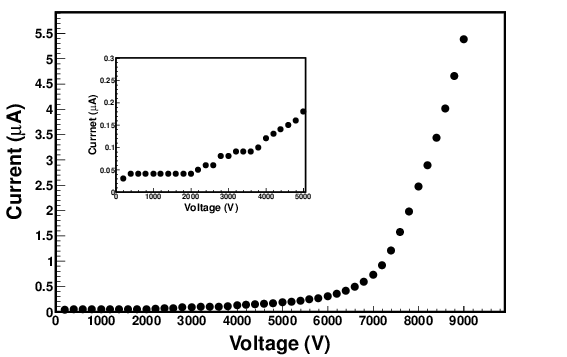}
\caption{\label{I-V} \small \sl I-V characteristics of the chamber. The figure in the inset shows the I-V characteristics at low
voltage region.}
\label{I-V} 
\end{center}
\end{figure}

\begin{figure}
\begin{center}
\includegraphics[height=5.0 cm, width=8.0 cm]{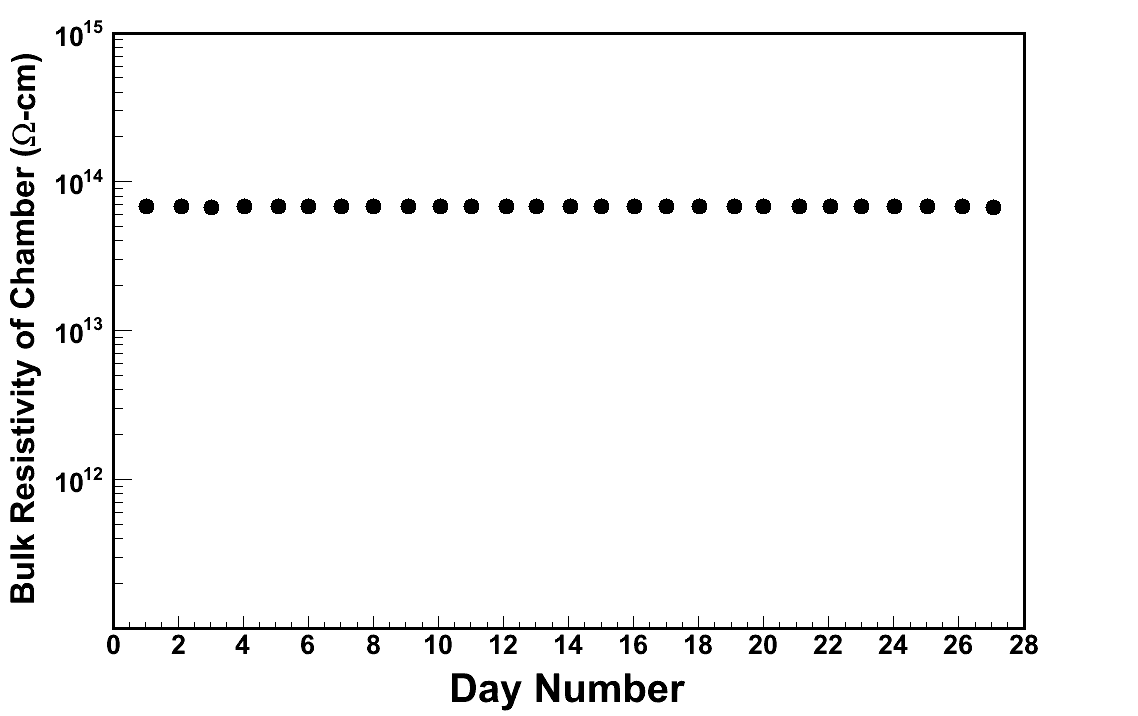}
\caption{\label{bulk_with_time} \small \sl [Color online] Variaton of bulk resistivity of the chamber with time.}
\label{bulk_with_time} 
\end{center}
\end{figure}

\subsection{Efficiency and noise rate}
We have studied the efficiency and the noise rate of the chamber at a signal threshold of -20 mV. 
Fig.~\ref{efficiency} shows the variation of efficiency with the total applied voltage showing a plateau of $>$95$\%$ 
above 8400V. The variation of efficiency, temperature and relative humidity during the
testing period is shown in Fig.~\ref{efficiency_time}. The noise rate variation as a function of the applied voltage is shown in Fig.~\ref{noise-rate}. 
During this test, the noise rate of the RPC has been found to be $\sim$0.75 Hz/$cm^{2}$ at 9000V. 
The noise rate is comparable to the value reported in \cite{12}.
 
\begin{figure}
\begin{center}
\includegraphics[height=5.0 cm, width=8.0 cm]{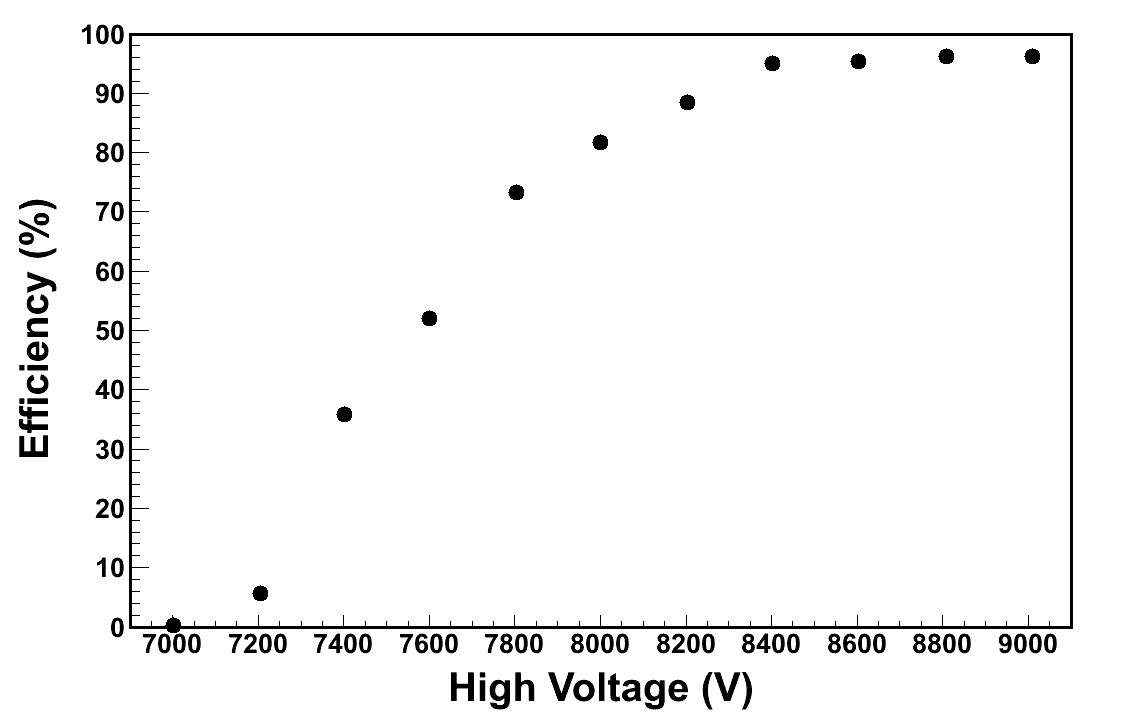}
\caption{\label{efficiency} \small \sl Efficiency of the chamber as a function of the applied voltage.
The error bars are within the marker size.}
\label{efficiency}
\end{center}
\end{figure}

\begin{figure}
\begin{center}
\includegraphics[height=5.0 cm, width=8.0 cm]{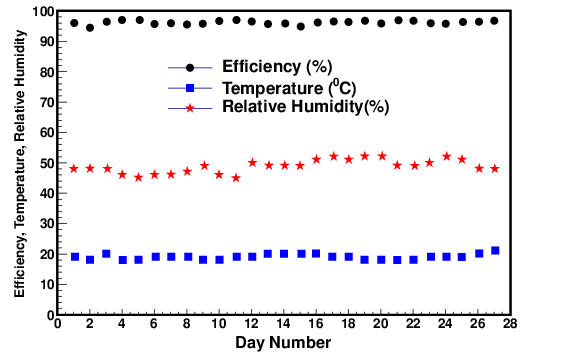}
\caption{\label{efficiency_time} \small \sl [Color online] variation of efficiency, temperature and rlative humidity with time.
The error bars of efficiencies are within the marker size.}
\label{efficiency_time}
\end{center}
\end{figure}

\begin{figure}
\begin{center}
\includegraphics[height=5.0 cm, width=8.0 cm]{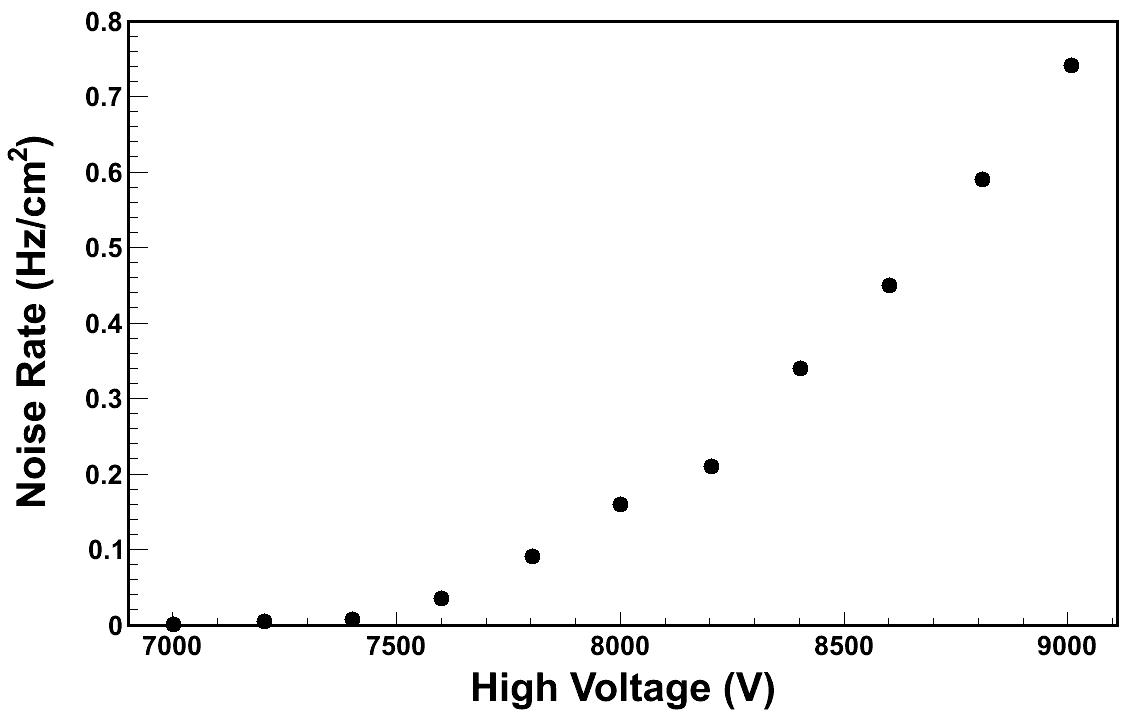}
\caption{\label{noise-rate} \small \sl Noise rate of the chamber as a function of the applied voltage.
The error bars are within the marker size.}
\label{noise-rate}
\end{center}
\end{figure}

\begin{figure}
\begin{center}
\includegraphics[height=5.0 cm, width=8.0 cm]{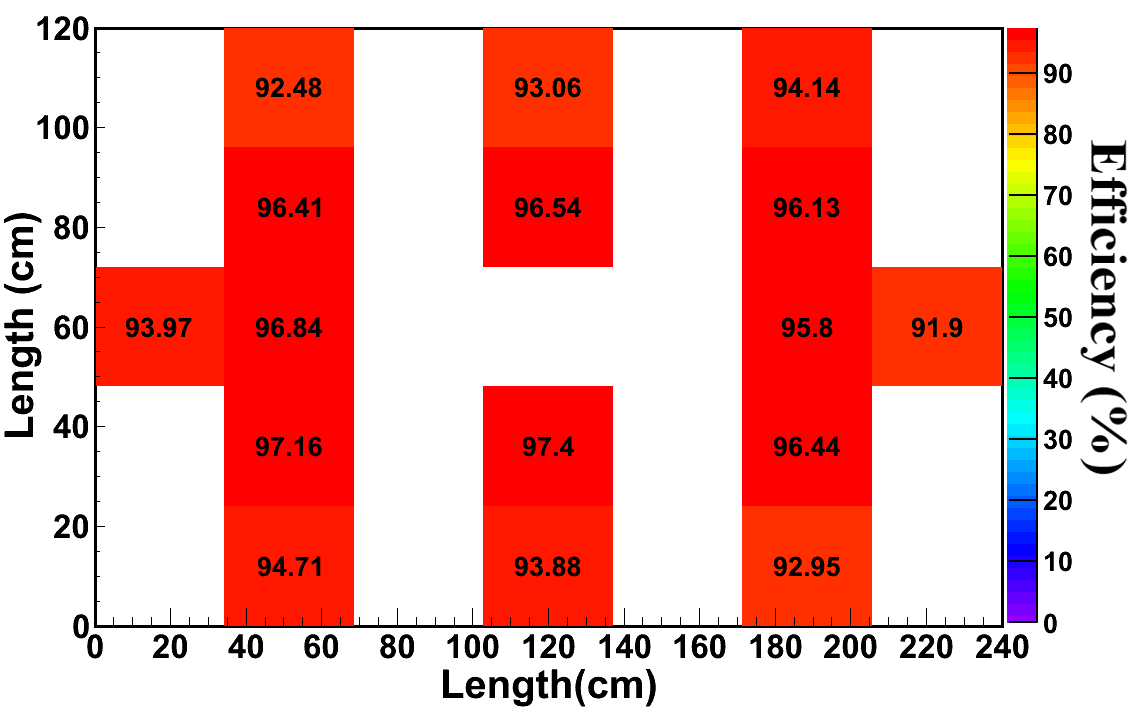}
\caption{\label{eff-locations} \small \sl [Color online] Locations on the RPC plane where efficiencies have been measured, with
		the measured efficiency values.}
\label{eff-locations}
\end{center}
\end{figure}

\begin{figure}
\begin{center}
\includegraphics[height=5.0 cm, width=8.0 cm]{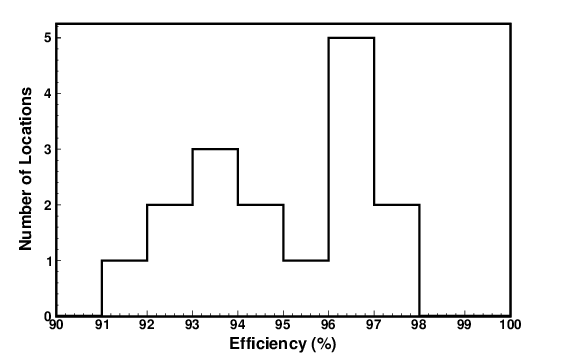}
\caption{\label{eff-uniformity} \small \sl Efficiency measurement at various locations on the RPC surface.}
\label{eff-uniformity}
\end{center}
\end{figure}

The efficiency of the chamber has been measured at 16 different locations of the detector at 9000V, 8 at the 
edges of the RPC
and 8 away from the edges. Fig.~\ref{eff-locations} shows these locations over the RPC surface with the measured efficiency 
values and Fig.~\ref{eff-uniformity}
shows the distribution of efficiency measured at these locations. The figure clearly shows two distinct groups, 
the edges of the RPC are relatively low efficient as the probability of
distortion of the electric field and the non-uniformity of gas-flow are higher at these regions. 
The average efficiency is found to be $>$95$\%$.

\subsection{Time resolution}

The time resolution of the RPC has been measured only at a central location of the chamber with the 16 channel 
PHILIPS SCIENTIFIC 7186 TDC module. Fig.~\ref{time-resolution-9kV} shows the uncorrected time spectra of the RPC at 9000 V.

\begin{figure}
\begin{center}
\includegraphics[height=5.0 cm, width=8.0 cm]{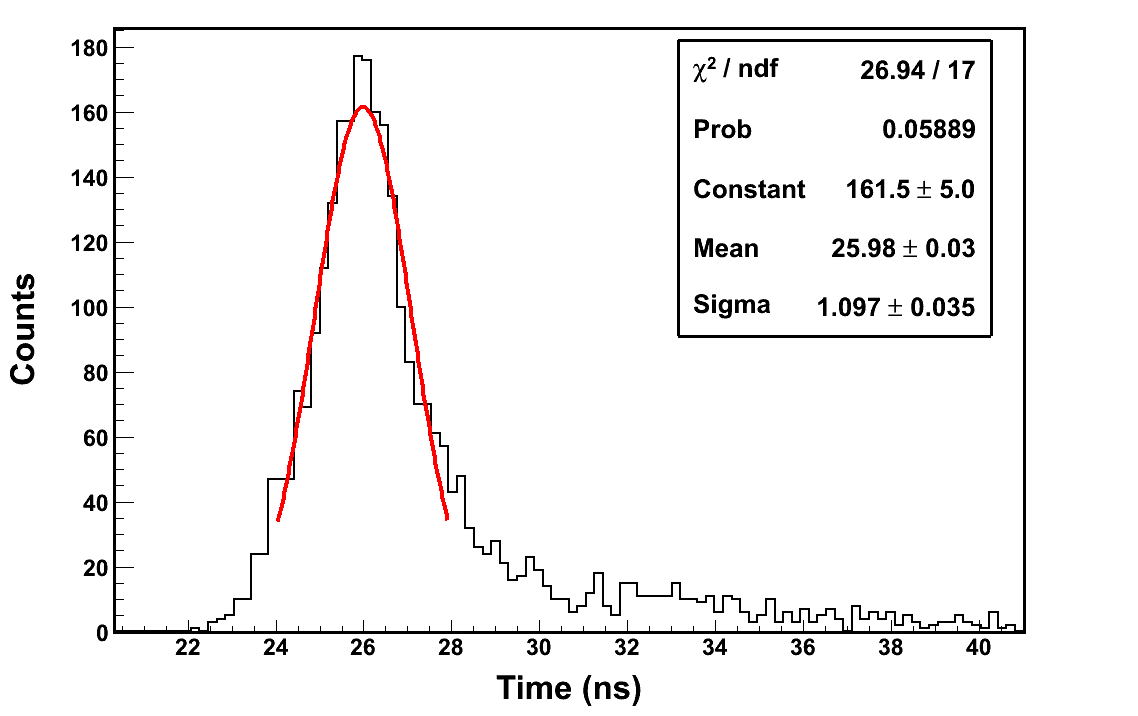}
\caption{\label{time-resolution-9kV} \small \sl[Color online] Raw TDC spectra of the RPC at 9000 V. The red curve shows the Gaussian fit.}
\label{time-resolution-9kV}
\end{center}
\end{figure}

The final RPC time resolution ($\sigma^{corrected}_{RPC}$) has been extracted from the Gaussian fit 
after removing the contribution of 
the three scintillators \cite{22}.
Fig.~\ref{time-resolution-HV} shows the variation of time resolution ($\sigma^{corrected}_{RPC}$) of the large RPC as a 
function of the applied voltage. The best value of the time
resolution has been found to be $\sim$0.83 ns at 9000V which is comparable to the
values reported in \cite{1},\cite{23}. 

\begin{figure}
\begin{center}
\includegraphics[height=5.0 cm, width=8.0 cm]{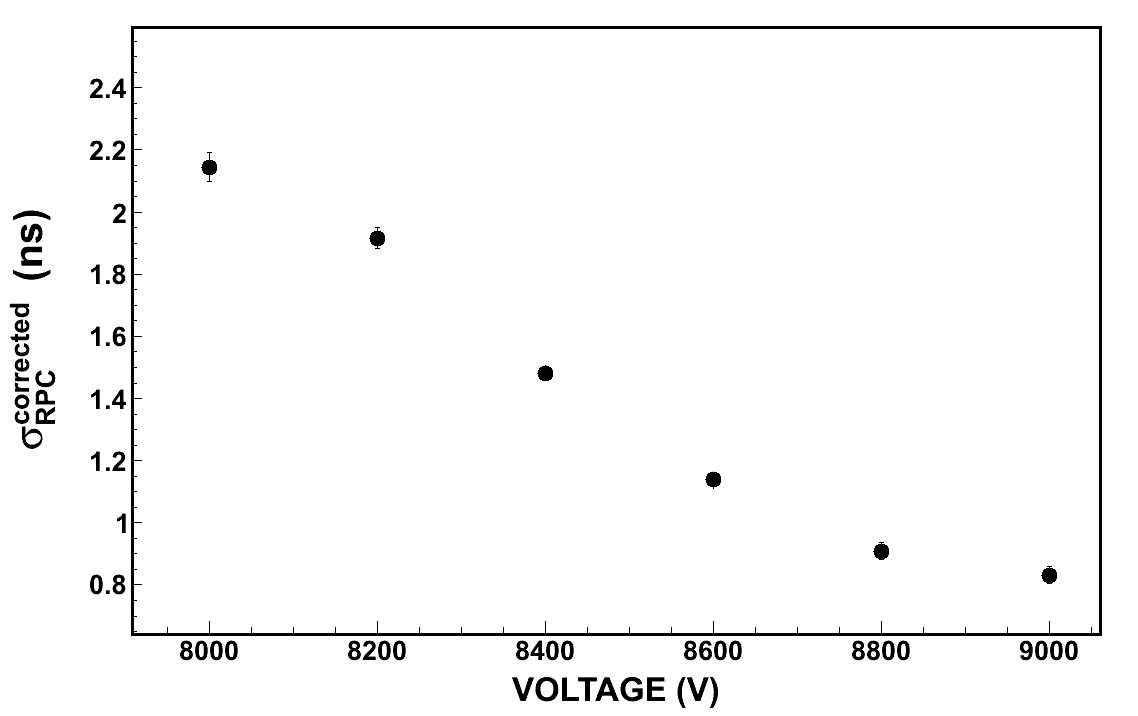}
\caption{\label{time-resolution-HV} \small \sl Time resolution ($\sigma^{corrected}_{RPC}$) as a function of the applied voltage.}
\label{time-resolution-HV}
\end{center}
\end{figure}

\section{ Conclusions}

Locally available P-301 OLTC grade HPL paper laminate has been characterised for its bulk and surface 
resistivities and its suitability for fabricating RPCs has been established.
A large oil-free HPL RPC of dimensions 240 cm $\times$ 120 cm $\times$ 0.2 cm,
made from the HPL samples, has been fabricated and tested. The 
challenges experienced during the fabrication of the large-sized RPC and the steps taken to overcome the issues,
the primary focus of this paper, have been highlighted.
The efficiency, noise rate and time resolution of the RPC
tested with cosmic rays in the streamer mode of operation at 9000V have been measured to be 
$>$95$\%$, $\sim$0.75 Hz/$cm^{2}$ and $\sim$0.83 ns respectively. The results obtained with this RPC make it suitable
to be used in large neutrino experiments.

\acknowledgments

The work is partially supported by the DAE-SRC award project fund of S. Chattopadhyay.
We are thankful to the INO collaborators for their encouragement.
We acknowledge the service rendered by Ganesh Das for making the glue samples and fabricating the detector. We also
acknowledge Ramanarayana Singaraju for his constant help throughout testing of this detector. 
We take this opportunity to thank Satyajit Saha and Y. P. Viyogi for their constant encouragement and all the fruitful
discussions and suggestions.

\end{document}